# A New Model of the Lifetime of Wireless Sensor Networks in Sea Water Communications

Abdelrahman Elleithy[1], Gonhsin Liu, Ali Elrashidi

Department of Computer Science and Engineering
University of Bridgeport,
Bridgeport, CT, USA
{aelleith@bridgeport.edu, gonhsin@bridgeport.edu, aelrashi@bridgeport.edu}

**Abstract.** In this paper we present a new model for the lifetime of wireless sensor networks used for sea water communications. The new model for power communications takes into consideration parameters such as power consumption for the active mode, power consumption for the sleep mode, power consumption for the transient mode, transmission period, transient mode duration, sleep mode duration, and active mode duration. The power communications model is incorporated in the life time model of wireless sensor networks. The life time model takes into consideration several parameters such as the total number of sensors, network size, percentage of sink nodes, location of sensors, the mobility of sensors, power consumption when nodes move and the power consumption of communications. The new model for power consumption in communications shows more accurate results about the lifetime of the sensor network in comparison with previously published results.

## 1 A Model for Lifetime of Wireless Sensors Networks

Wireless sensors have received increased attention in the past years due to their popularity and cost effectiveness when they are used in harsh environments. They

1 The corresponding author.

have been used in many applications including military applications, environmental applications, health applications, and home applications. Although they are very cost effective and easily deployed in harsh environments, they are limited by the power available through their life cycle. Sensors are usually deployed with limited power which is depleted over their life cycle. Once their power is depleted, the sensors become dead and they are no more useful. An evaluation of the life cycle of a wireless sensor network is very essential to estimate how long a network can live and when the network and its sensors might be replaced or recharged if possible.

In this section we present a model for the lifetime of Wireless sensor networks based on a paper by [1]. The model takes different parameters that are used in literature. The following parameters are considered:

1. The time until the first sensor is drained of its energy [2];
2. The time until the first cluster head is drained of its energy [3];
3. The time there is at least a certain fraction β of surviving nodes in the network [4];
4. The time until all nodes have been drained of their energy [5];
5. K-coverage: the time the area of interest is covered by at least k nodes [6];
6. 100% coverage

a. The time each target is covered by at least one node [7] ;

b. The time the whole area is covered by at least one node [8] ;

7. α-coverage

a. The accumulated time during which at least α portion of the region is covered by at least one node [9];

b. The time until the coverage drops below a predefined threshold α (until last drop below threshold) [10] ;

c The continuous operational time of the system before either the coverage or delivery ratio first drops below a predefined threshold [11];

8. The number of successful data-gathering trips [12] ;
9. The number of total transmitted messages [13];
10. The percentage of nodes that have a path to the base station [11];
11. Expectation of the entire interval during which the probability of guaranteeing connectivity and k-coverage simultaneously is at least α [6];

12. The time until connectivity or coverage are lost [14];
13. The time until the network no longer provides an acceptable event detection ratio [5];
14. The time period during which the network continuously satisfies the application requirement [15];
15. The minimum of $t_1$, $t_2$, and $t_3$ with $t_1$: time for cardinality of largest connected component of communication graph to drop below $c_1 \times n(t)$, $t_2$: time for n(t) to drop below $c_2 \times n$, $t_3$: time for the covered volume to drop below $c_3 \times l^d$ [16].

## 2 Parameters Used In the Model

In this section we address parameters that were introduced in literature that can be used in a complete model for a wireless sensors networks lifetime. In earlier version of this model was introduced by Elleithy and Liu [17]. The following parameters are introduced:

1. The total number of available sensors
2. The set of all nodes those that are alive at time t
3. The set of nodes those that are active at time t
4. The set of nodes those that are active at any time in the time interval $[t - \Delta t, t]$
5. The set of sink nodes or base stations B(t) is defined to be a subset of the existing nodes $S^Y$
6. The ability of nodes $m_l$ and $m_n$ to communicate at a time t
7. The ability of two nodes to communicate in the time interval $[t - \Delta t, t]$ such that the links between consecutive hops become available successively within the time interval (support for delay tolerant networking)
8. The set of target points to be sensed by the network
9. The area that is covered by all sensors of a certain type y, at a time t.

## 3 Definition of Network Lifetime

There are two network lifetime metrics that introduced in literature based on definitions in the previous sections. Both metrics depict the network lifetime in seconds. The metrics probably become most expressive when used together.

(1) The first metric gives the accumulated network lifetime $Z_a$ as the sum of all times that ζ (t) is fulfilled, stopping only when the criterion is not fulfilled for longer than $\Delta t_{sd}$ seconds.

(2) The second metric, the total network lifetime $Z_t$, gives the first point in time when the liveliness criterion is lost for a longer period than the service disruption tolerance $\Delta t_{sd}$.

## 4 A New Model of Power Consumption in Communications of Wireless Sensor Networks

In this section we present a new model for the power consumption of the communications in wireless sensor networks. The model is based on the work of Cui *et. al.* [18].

### 4.1 Energy Consumed in Communications

In this model, the total energy consumed in communications is given by Equation (1):

$$E = P_{on}T_{on} + P_{sp}T_{sp} + P_{tr}T_{tr} = (P_t + P_{c0})T_{on} + P_{sp}T_{sp} + P_{tr}T_{tr} \quad (1)$$

The following parameters are used in the calculation of Equation (1):

1. $P_{on}$ power consumption value for the active mode
2. $P_{sp}$ power consumption value for the sleep mode
3. $P_{tr}$ power consumption value for the transient mode
4. the transmission period: $T$ is given by $T = T_{tr} + T_{on} + T_{sp}$.

5. $T_{\text{tr}}$ is the transient mode duration, which is equal to the frequency synthesizer settling time (the start-up process of the mixer and PA is fast enough to be neglected),
6. $T_{\text{sp}}$ is the sleep mode duration,
7. $T_{\text{on}}$ is the active mode time for the transceiver such that $T_{\text{on}} \le T$ , where $T_{\text{on}}$ is a parameter to optimize,
8. The active mode power $P_{\text{on}}$ comprises the transmission signal power $P_t$ and the circuit power consumption $P_{c0}$ in the whole signal path.
9. $P_{c0}$ consists of the mixer power consumption $P_{\text{mix}}$, the frequency synthesizer power consumption $P_{\text{syn}}$, the LNA power consumption $P_{\text{LNA}}$, the active filter power consumption $P_{\text{filt}}$ at the transmitter, the active filter power consumption $P_{\text{filr}}$ at the receiver, the IFA power consumption $P_{\text{IFA}}$, the DAC power consumption $P_{\text{DAC}}$, the ADC power consumption $P_{\text{ADC}}$, and the PA power consumption $P_{\text{amp}}$
10. $P_{\text{amp}} = \alpha P_t$ and $\alpha = \xi/\eta - 1$ with $\eta$ the drain efficiency [11] of the RF PA and $\xi$ the peak-average ratio (PAR), which is dependent on the modulation scheme and the associated constellation size.
11. Since $P_{\text{on}} = \max\{P_{\text{on}}, P_{\text{tr}}, P_{\text{sp}}\}$, the peak-power constraints are given by:

$P_{\text{ont}} = P_t + P_{\text{amp}} + P_{ct} = (1+\alpha)P_t + P_{ct} \le P_{\text{maxt}}$

$P_{\text{onr}} = P_{cr} \le P_{\text{maxr}}$

12. $P_{ct} = P_{\text{mix}} + P_{\text{syn}} + P_{\text{filt}} + P_{\text{DAC}}$ and $P_{cr} = P_{\text{mix}} + P_{\text{syn}} + P_{\text{LNA}} + P_{\text{filr}} + P_{\text{IFA}} + P_{\text{ADC}}$ denote the circuit power consumption (excluding the PA power consumption) in the active mode at the transmitter and the receiver, respectively.
13. The start-up time for other circuit blocks is negligible compared to that of the frequency synthesizers.
14. Given (1) and (2), and the fact that $P_{\text{sp}} = 0$and $P_{\text{tr}} \approx 2P_{\text{syn}}$, the energy consumption per information bit $E_a = E/L$ is given by
15. $E_a = [(1 + \alpha)P_t T_{\text{on}} + P_c T_{\text{on}} + P_{\text{tr}} T_{\text{tr}}]/L \approx [(1 + \alpha)E_t + P_c T_{\text{on}} + 2P_{\text{syn}} T_{\text{tr}}]/L$
16. $B_e = L/(BT_{\text{on}})$ (in bits per second per hertz).

### 4.2 Under water Signal Propagation

The signal propagation in water depends on the path loss in water. Received power as a function of transmitted signal, path loss and antenna gain at the receiver end is given from Friis equation as shown in Equation 2 [19].

$$P_{rec}(dBm) = P_i(dBm) + G_t(dB) + G_r(dB) - L_{pathloss}(dB) \qquad (2)$$

where $P_t$ is the transmit power, $G_r$ and $G_t$ are the gains of the receiver and transmitter antenna, $L_{Pathloss}$ is the path loss in water.

The path loss is shown in Equation 3 [20].

$$L_{pathloss}(dB) = L_0(dB) + L_w(dB) + L_{att}(dB) \qquad (3)$$

$L_0$ is the path loss in air and given by:

$$L_0(dB) = 20\log\left(\frac{4\pi df}{c}\right) \qquad (4)$$

where $d$ is the distance between transmitter and receiver in meter, $f$ is the operating frequency in Hertz and $c$ is the velocity of light in air in meter per second.

$L_w(dB)$ is the path loss due to changing in medium and given by [19]:

$$L_w(dB) = 20\log\left(\frac{\lambda_0}{\lambda}\right) \qquad (5)$$

where $\lambda_0$ is the signal wavelength in air and calculated ($\lambda_0 = c/f$) and $\lambda$ is the wave factor and given by ($\lambda = 2\pi/\beta$) and $\beta$ is the phase shifting constant and calculated as shown in Equation 6.

$$\beta = \omega\sqrt{\frac{\mu\varepsilon'}{2}\left(\sqrt{1 + \left(\frac{\varepsilon''}{\varepsilon'}\right)^2} + 1\right)} \qquad (6)$$

where $\varepsilon'$ and $\varepsilon''$ are the real and imaginary parts of the complex dielectric constant given by ($\varepsilon = \varepsilon' - j\varepsilon''$).

$L_{att}(dB)$ is the path loss due to attenuation in medium and given by:

$$L_{att}(dB) = 10\log(e^{-2\alpha d}) \qquad (7)$$

where $\alpha$ is the attenuation constant and calculated as shown in Equation 8:

$$\alpha = \omega\sqrt{\frac{\mu\square'}{2}(\sqrt{1+\left(\frac{\square''}{\square'}\right)^{2}}-1)} \qquad (8)$$

The reflection from the surface and bottom depends on reflection coefficient at the interface between water and air and between water and sand. The reflection coefficient is given by Equation 9 [20].

$$\Gamma = \frac{\rho_2 v_2 - \rho_1 v_1}{\rho_2 v_2 + \rho_1 v_1} \qquad (9)$$

where $\rho_1$ and $\rho_2$ are the density of the first and second medium respectively and $v_1$ and $v_2$ are the wave velocity in both mediums.

The reflection loss from the surface and from the bottom is $L_{ref}$ and shown in Equation 10.

$$L_{ref} = -V(dB) = -10\log(V) \qquad (10)$$

where is calculated as shown below:

$$V^2 = 1 + \left(|\Gamma|e^{-\alpha\Delta(r)}\right)^2 - 2|\Gamma|e^{-\alpha\Delta(r)} \times \cos\left(\pi - \left(\phi - \frac{2\pi}{\lambda}\Delta(r)\right)\right) \qquad (11)$$

where $r$ is the reflected path length, $|\Gamma|$and $\phi$ are the amplitude and phase of the reflection coefficient respectively and $\Delta(r)$ is the difference between $r$ and $d$.

where $r$ can be calculated as follow:

$$r = 2\sqrt{H^2 + (\frac{d}{2})^2} \qquad (12)$$

### 4.3 Simulation Results

The following parameters values are used in the simulation carried in this paper:

| | | |
|---|---|---|
| k = 3.5 | $G_1$ = 1000 Watt | B = 10 KHz |
| $P_{mix}$ = 30.3 mW | $P_{LNA}$ = 20 mW | $P_{maxt}$ = 250 mW |
| $T_{tr}$ = 5 micro-second | T = 0.1 second | drainEfficiency = 0.35 |
| psdensity = 5.0119*$10^{-12}$ Watt per Hz | | L = 2 Kbit |
| $P_{syn}$ = 50 mW | $P_{IFA}$ = 3 mW | $P_{filt}$ = 2.5 mW |
| $P_{filr}$ = 2.5 mW | $M_l$ = 10000 Watt | $P_b$ = 0.001 |
| $V_{dd}$ = 3 Volt | $L_{min}$ = 0.5 micrometer | $n_1$ = 10 |
| $n_2$ = 10 | $f_{cor}$ = 1 MHz | $I_0$ = 10 microampere |
| Cp = 1 picofarad | beta = 1; | |

In this section we show the results of the power consumption in communications where the total energy per bit is calculated versus time spent on the On mode for different parameters.

### 4.3.1 Energy Consumption for Sea Water

In Figure 1, the Total Energy Consumption per bit versus Time Spent in the On mode / Total Time for different distances at a packet size of 2000 bits. The figure shows the energy consumption from 1 meter to 5 meters. As the distance of transmission increases, the total energy per bit increases. Also, as the time Spent in the On mode / Total Time increases, the energy per bit decreases.

In Figure 2, the Total Energy Consumption per bit versus Time Spent in the On mode / Total Time for different packet size for a path loss exponent of 3.5 and a distance of 3 meters. As the packet size increases, the total energy per bit increases. Also, as the time Spent in the On mode / Total Time increases, the energy per bit decreases.

In Figure 3, the Total Energy Consumption per bit versus Time Spent in the On mode / Total Time for different path loss exponent (K) at a packet size of 2000 bits and a distance of 3 meters. As the path loss exponent increases, the total energy per bit increases. Also, as the time Spent in the On mode / Total Time increases, the energy per bit decreases.

In Figure 4, the Total Energy Consumption per bit versus Time Spent in the On mode / Total Time for different bandwidth at a packet size of 2000 bits and a distance of 3 meters. As the bandwidth decreases, the total energy per bit increases. Also, as the time Spent in the On mode / Total Time increases, the energy per bit decreases.

In Figure 5, the Total Energy Consumption per bit versus Time Spent in the On mode / Total Time for different drain efficiency at a packet size of 2000 bits and a distance of 3 meters. As the drain efficiency decreases, the total energy per bit increases. Also, as the time Spent in the On mode / Total Time increases, the energy per bit decreases.

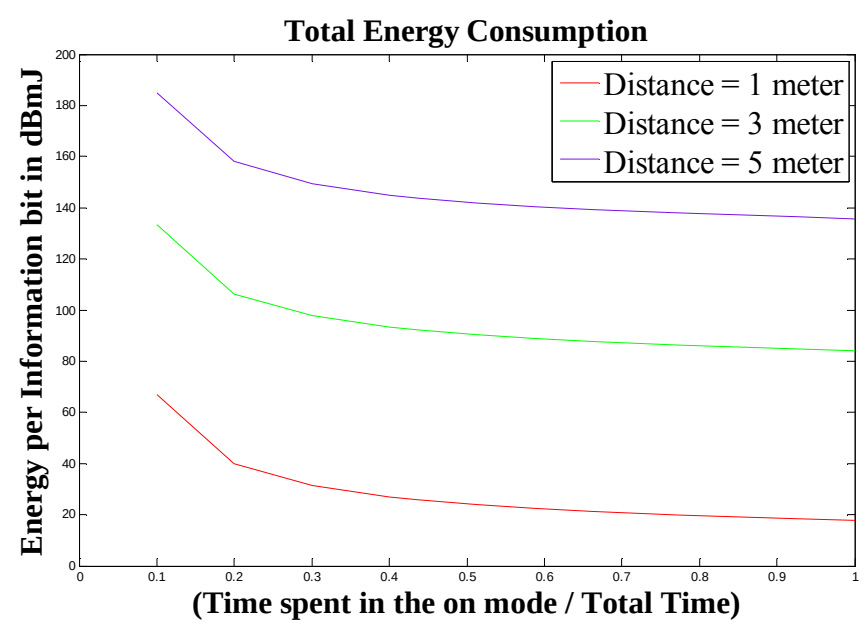

**Fig. 1.** Total Energy Consumption per bit versus Time Spent in the On mode / Total Time for different distances at a packet size of 2000 bits.

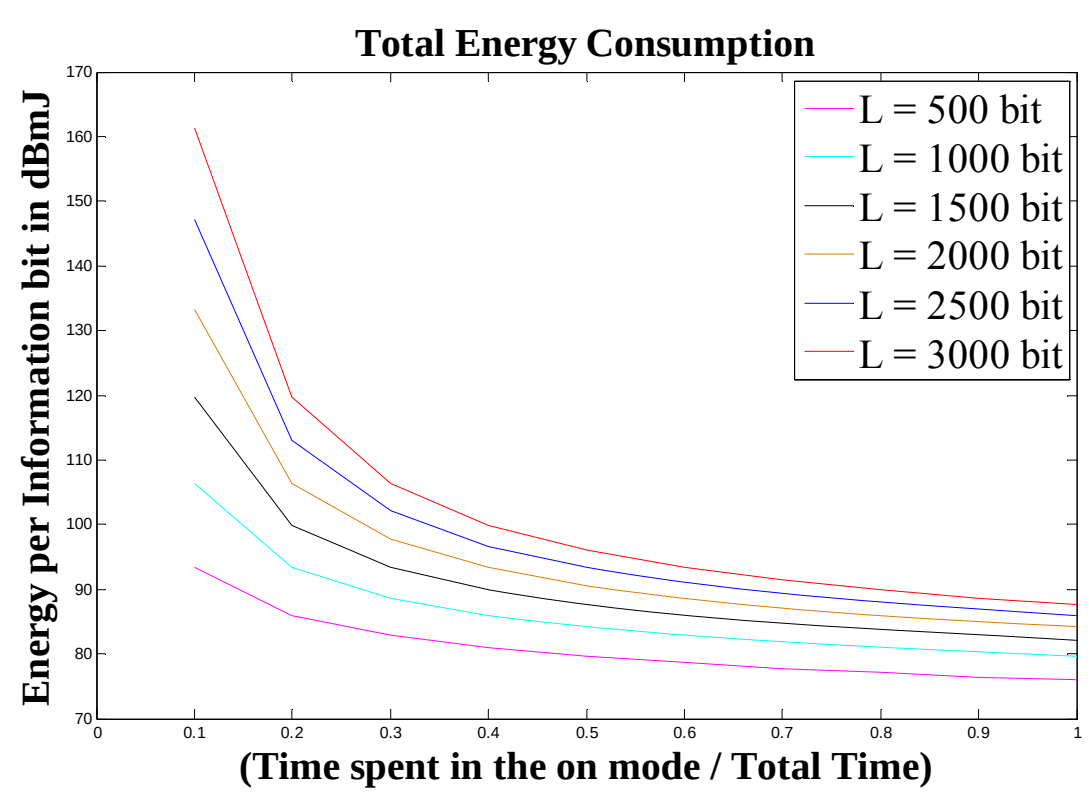

**Fig. 2.** Total Energy Consumption per bit versus Time spent in the On mode / Total Time for different packet sizes at a distance of 3 meter.

## 5 Simulation of the Lifetime of Wsn

In this section we present the following model for simulating the life time in a WSN. Although the Matlab code is developed with the following default parameters, it can be modified for other values.

Assumptions:

1. Total number of sensors: Input parameter
2. Network size: Input parameter in meters
3. Percentage of sink sensors: Input parameter (between 0 and 1)
4. Location of sensors: Randomly generated over the network
5. Initial power per sensor: Random between 0 and 100 units
6. Movement of sensors:
   a. sensors can move in the x direction for a random value between -5 and +5
   b. sensors can move in the y direction for a random value between -5 and +5
   c. The total value a senor moves is: $d = \sqrt{x^2 + y^2}$
7. Power Consumption:
   a. Communications: As given by equation 1
   b. Movement: d unit for the moving sensor as calculated in 6
8. Stopping criteria (we consider the network dead if one of the following conditions satisfied):
   a. Percentage of available power to total power: less than 25 %
   b. Percentage of alive sensors to total sensors: less than 25 %
   c. Percentage of alive sink sensors to total sink sensors: less than 5 %

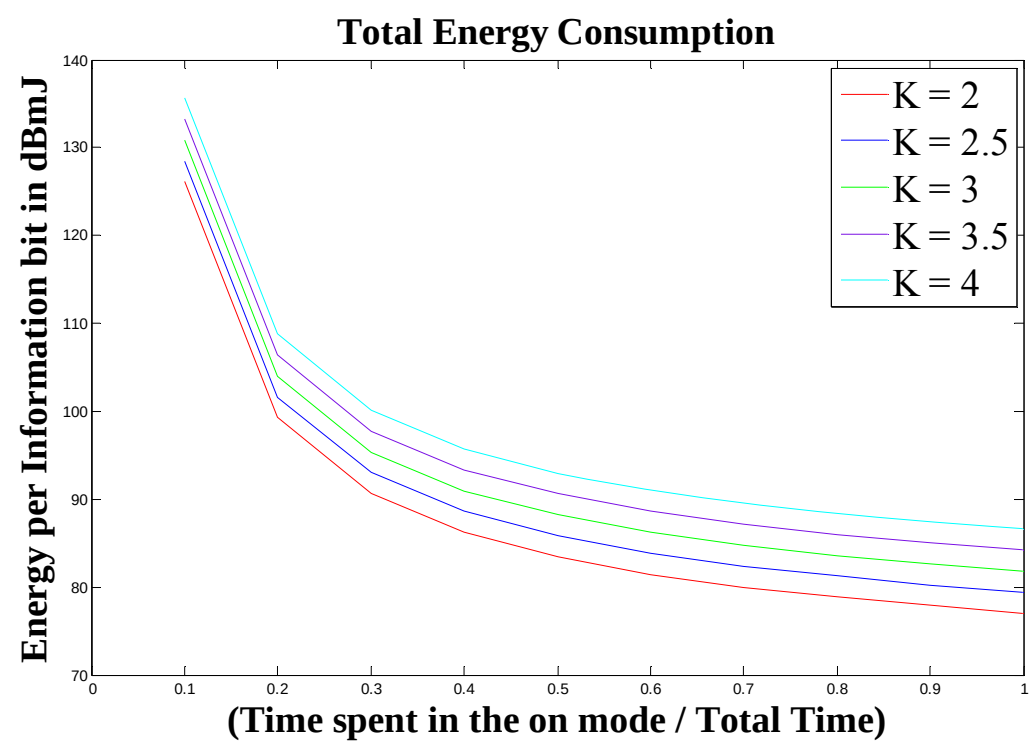

**Fig. 3.** Total Energy Consumption per bit versus Time Spent in the On mode / Total Time for different path loss exponent (k) at a distance of 3 meter.

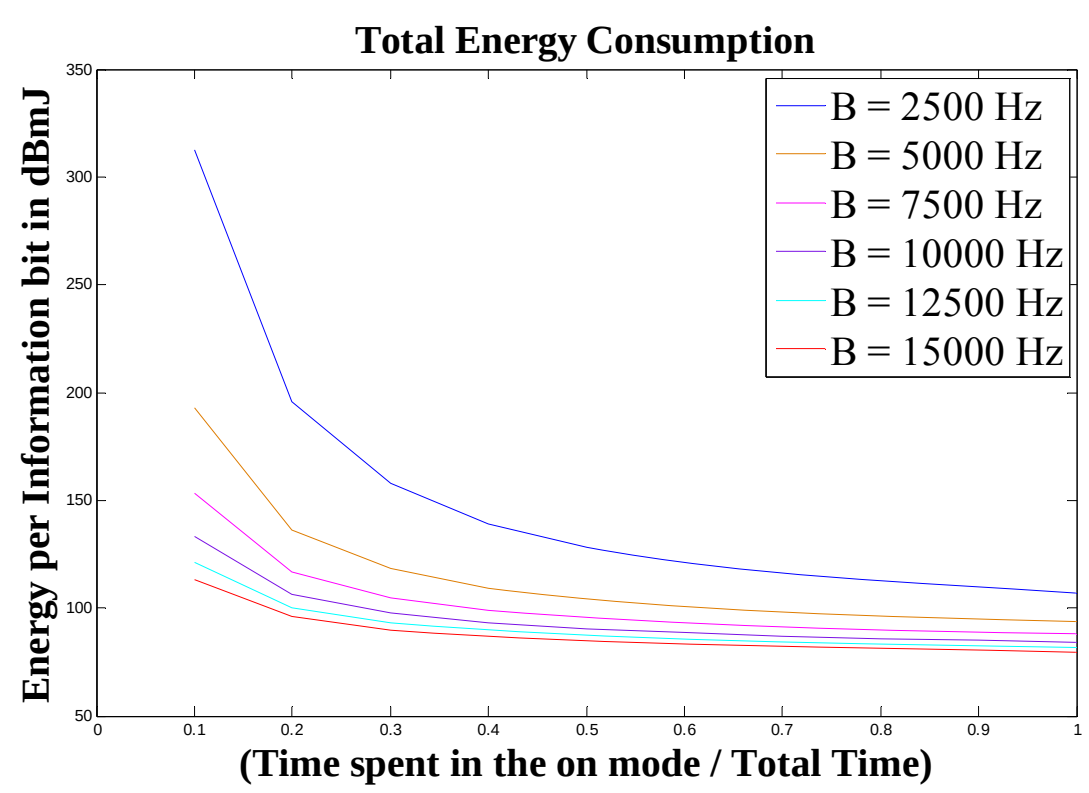

**Fig.4.** Total Energy Consumption per bit versus Time Spent in the On mode / Total Time for different bandwidth at a distance of 3 meter.

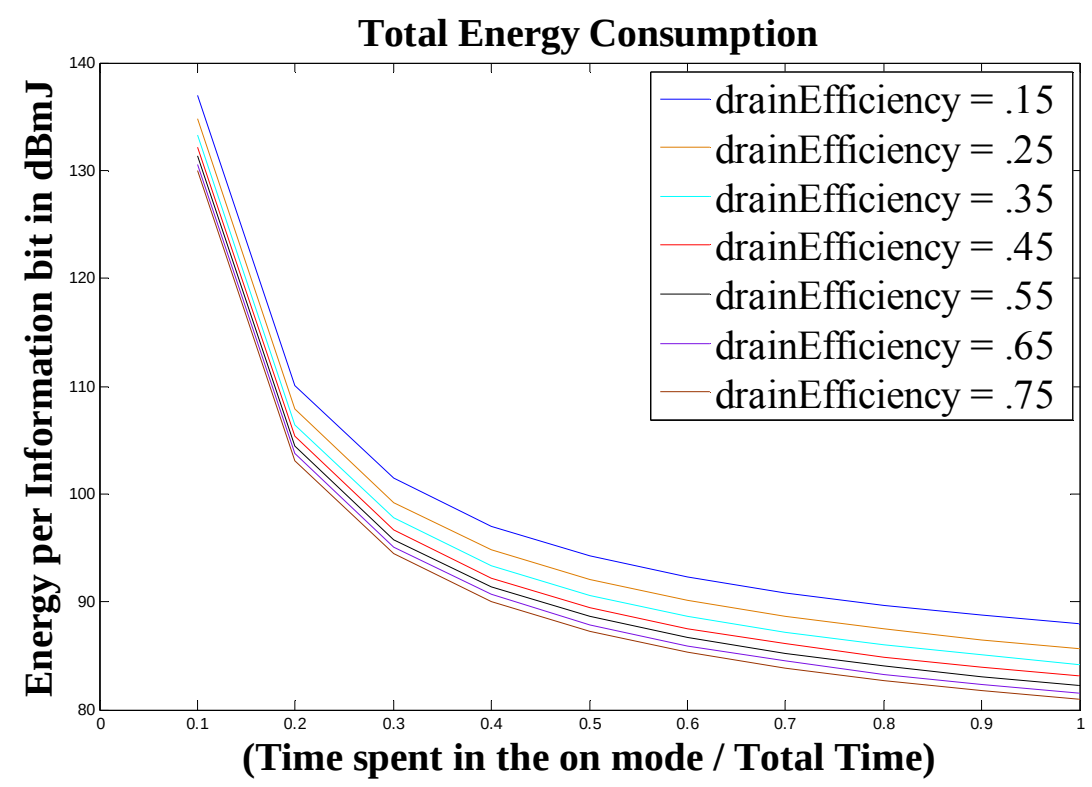

**Fig. 5.** Total Energy Consumption per bit versus Time Spent in the On mode / Total Time for different drain Efficiency at a distance of 3 meter.

### 5.1. Scenario Number 1

In this scenario the following parameters are considered:

1. Total number of sensors: 150
2. Network size: 3000 m X 3000 m
3. Percentage of sink sensors: 15.6 %
4. Maximum Communications power: 12.98 units
5. Packet size (k): 32

Figure 6 shows the status of Initial and final network. Figures 7 to 10 show different network performance parameters over the life cycle of the network. In this scenario the network was considered dead because after 22 cycles the following status is reached:

1. Percentage of available power to total power:

   1785 / 7716 = 23 %

2. Percentage of alive sensors to total sensors:

   73 / 150 = 49 %

3. Percentage of alive sink sensors to total sink sensors:

5 / 20 = 25 %

The network is considered dead because condition number (1) is satisfied where the percentage of available power to total power is 23 % which is less than the 25 % criteria.

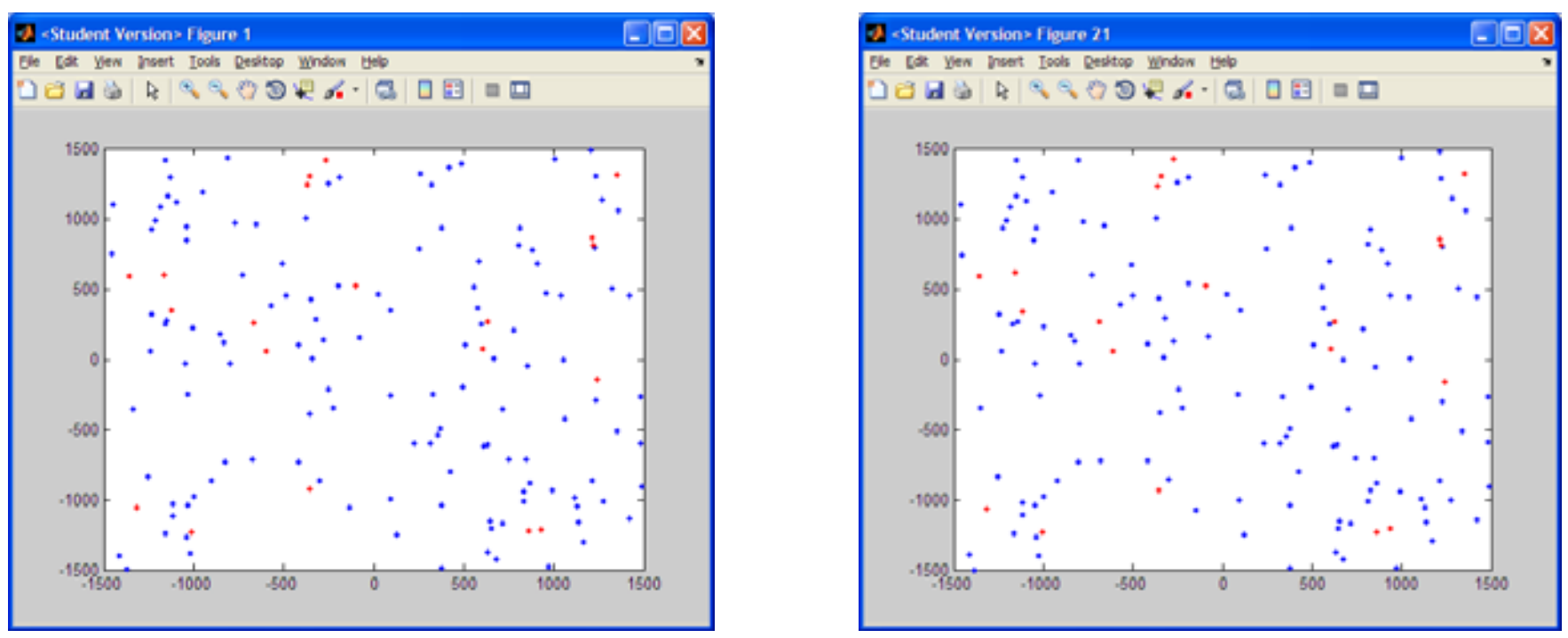

(a) Initial Network (b) Final Network

Note: sink nodes are red

**Fig. 6.** Status of Initial and final network.

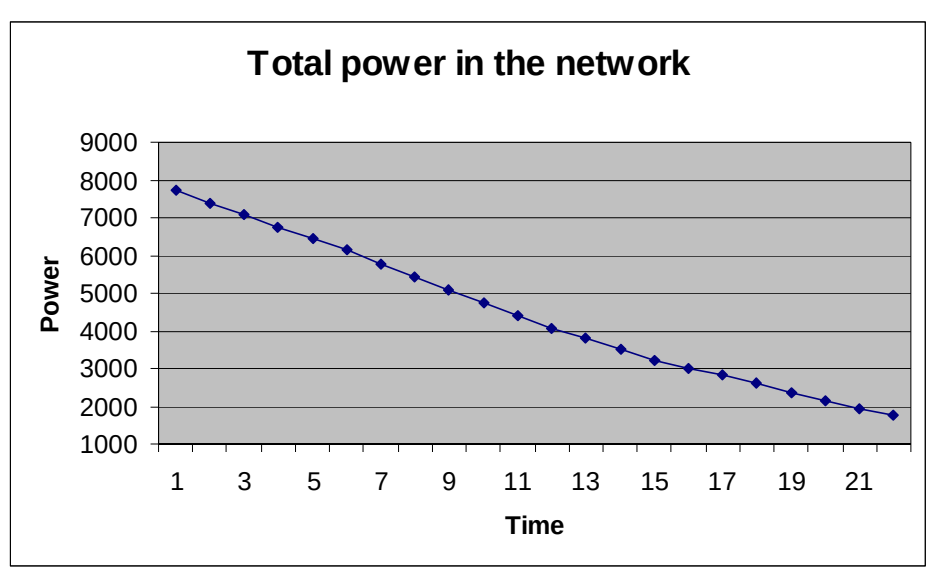

**Fig. 7.** Total Power in the network over the life cycle

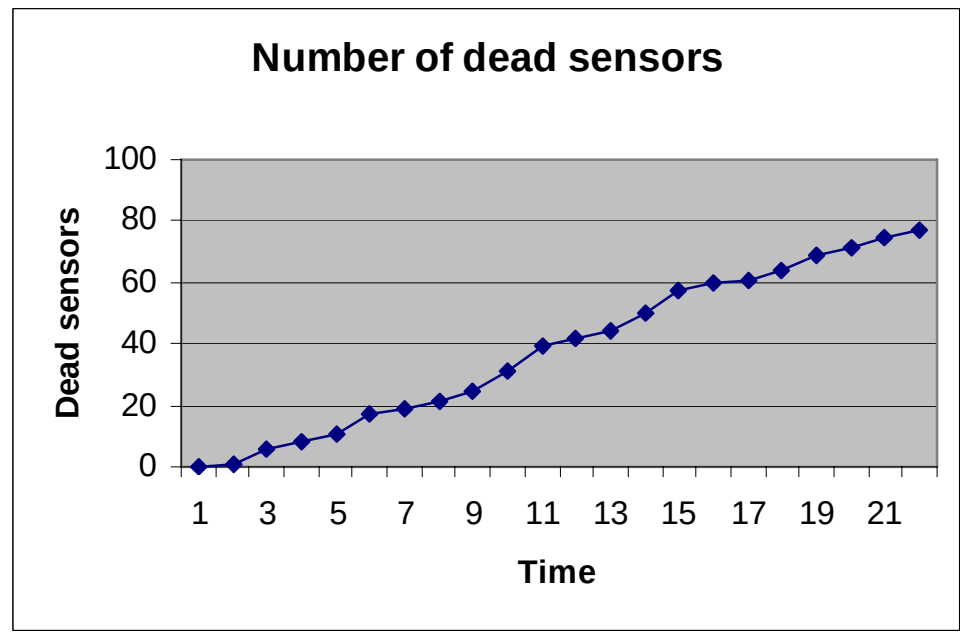

**Fig. 8.** Numbers of dead sensors over the life cycle

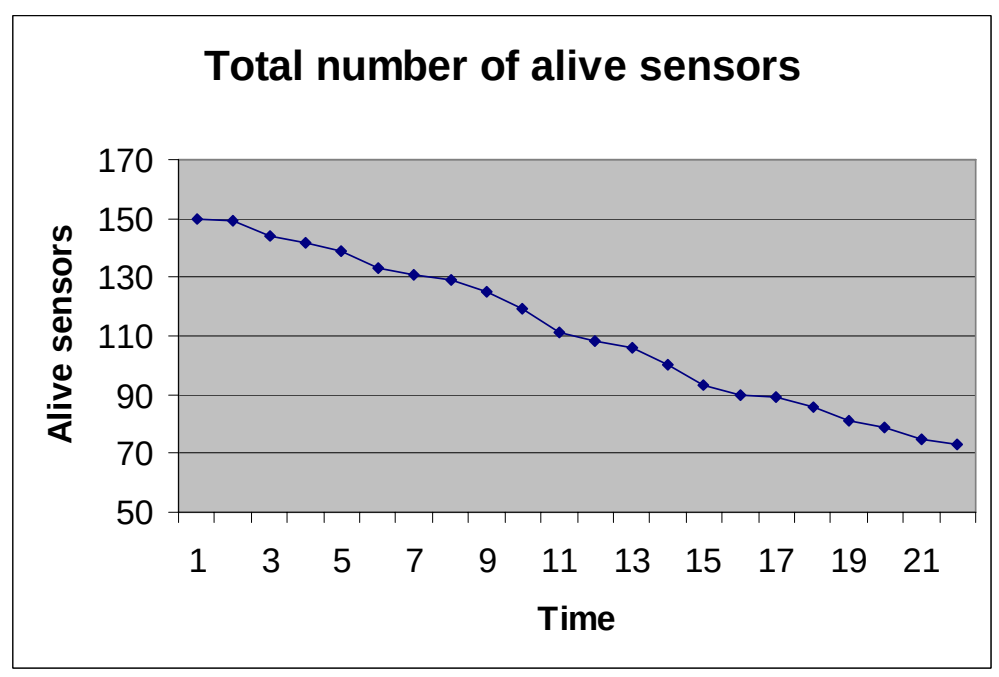

**Fig. 9.** Total number of alive sensors over the life cycle

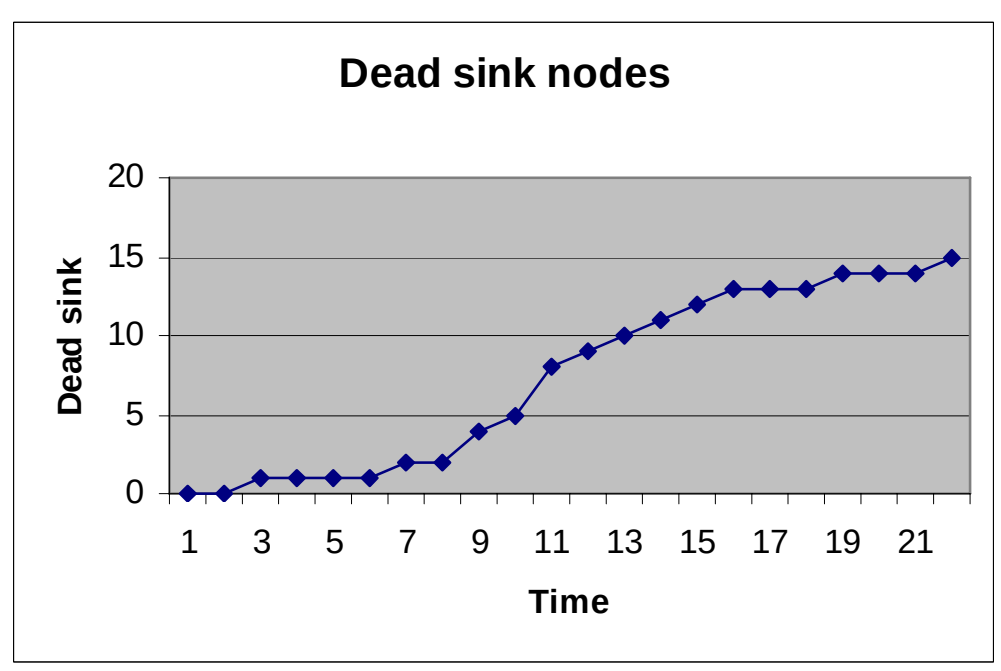

**Fig. 10.** Total number of dead sinks nodes over the life cycle

### 5.2. Scenario Number 2

In this scenario the following parameters are considered:

1. Total number of sensors: 150
2. Network size: 3000 m X 3000 m
3. Percentage of sink sensors: 15.6 %
4. Maximum Communications power: 11.73 units
5. Packet size (k): 32

Figure 11 shows the status of Initial and final network. Figures 12 to 15 show different network performance parameters over the life cycle of the network. In this scenario the network was considered dead because after 22 cycles the following status is reached:

1. Percentage of available power to total power: 1824 / 7689= 23.7 %
2. Percentage of alive sensors to total sensors: 75 / 150 = 50%
3. Percentage of alive sink sensors to total sink sensors: 8 / 21 =38 %

The network is considered dead because condition number (1) is satisfied where the percentage of available power to total power is 23.7 % which is less than the 25 % criteria.

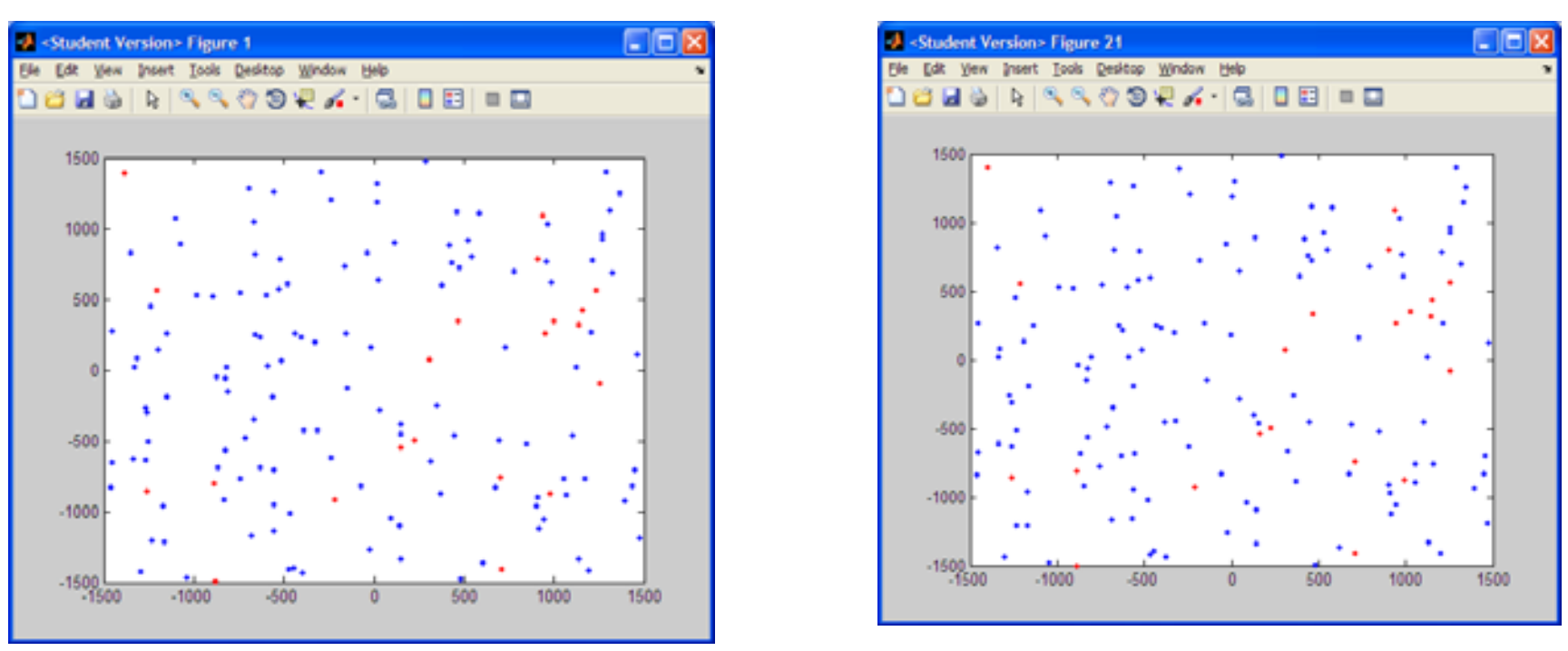

(a) Initial Network (b) Final Network

Note: sink nodes are red.

**Fig. 11.** shows the status of Initial and final network.

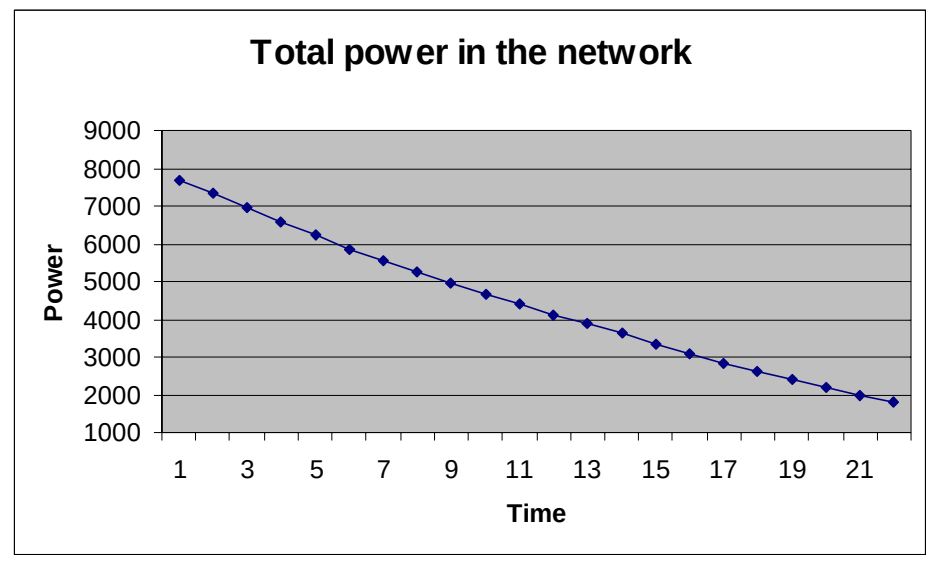

**Fig. 12.** Total Power in the network over the life cycle

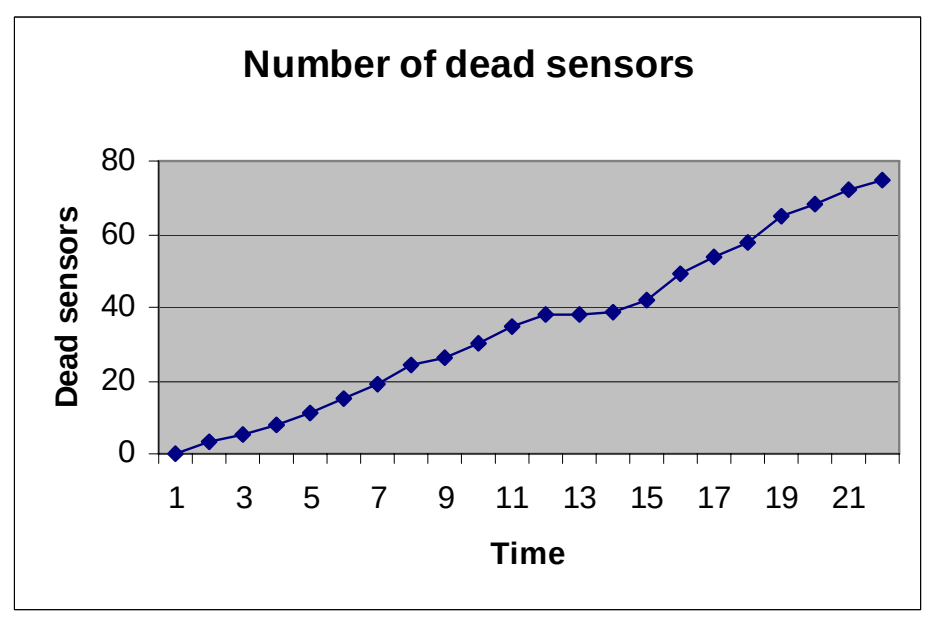

**Fig. 13.** Numbers of dead sensors over the life cycle

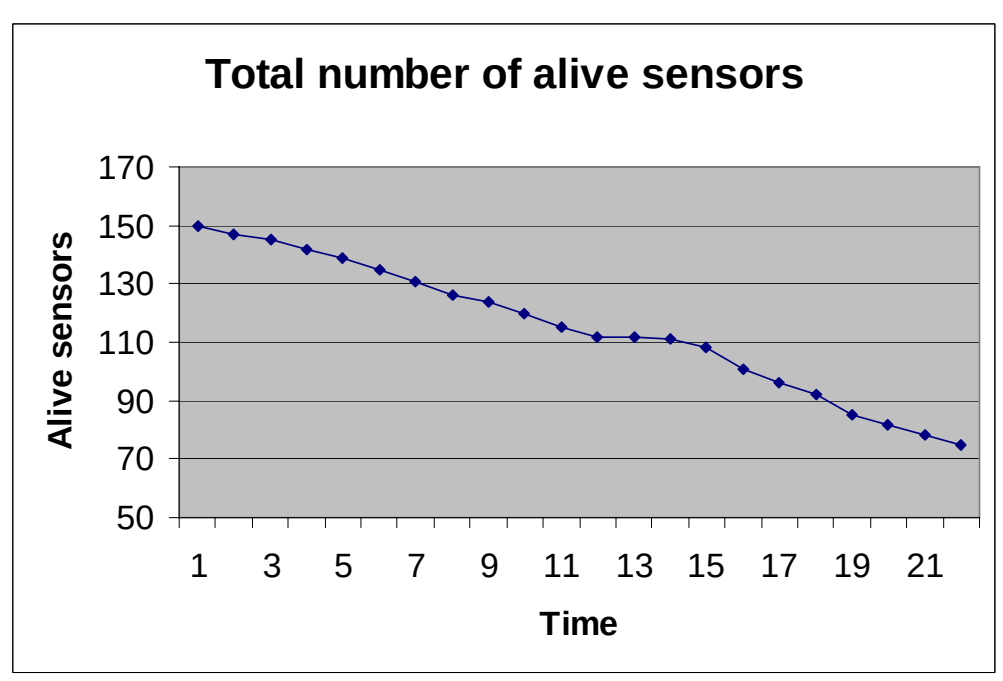

**Fig. 14.** Total number of alive sensors over the life cycle

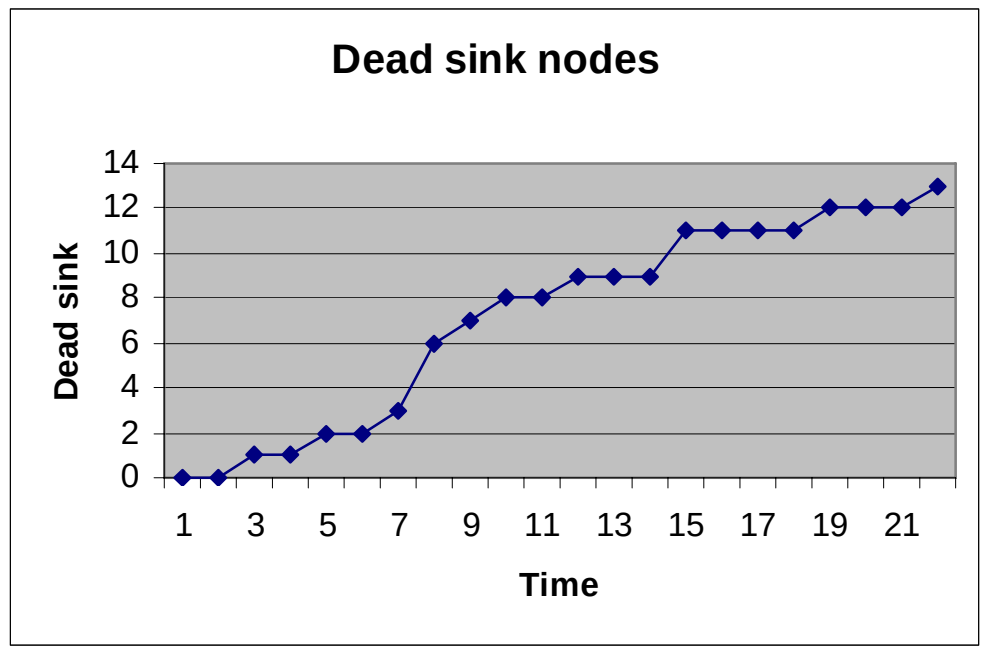

**Fig. 15.** Total number of dead sinks nodes over the life cycle

## 6 Conclusions

Although wireless sensors networks are very popular, their usage is restricted due to the limited power available through their life cycle. Once their power is depleted, sensors might be replaced or recharged if possible. A model to estimate the lifetime of wireless sensors networks is necessary to help the designers of the network to design their network by adjusting important parameters such as initial power, number of sensors, number of sink sensors, etc.

In this paper we present a model for the power consumed in communications of the wireless sensors. The presented model for power communications takes into consideration parameters such as power consumption for the active mode, power consumption for the sleep mode, power consumption for the transient mode, transmission period, transient mode duration, sleep mode duration, and active mode duration.

In order to examine the validity of our model, we have tested it for many lifetime scenarios. In this paper we are presenting five of these scenarios. The following parameters are used: total number of sensors, network size as defined by its width and length, and the percentage of sink sensors. In each scenario, we have evaluated both the total power in the network over the life cycle, number of dead sensors over the life cycle, total number of alive sensors over the life cycle, and number of dead sinks nodes over the life cycle.

The results presented in this paper show the importance of such a simulator from the designer perspective. The model can be used as a design tool as well as a research tool to evaluate the network performance. In the future, we would expect to extend the work presented in this model to include other parameters and modes of operations for underwater and underground wireless sensor networks.